\documentclass[a4paper,11pt]{article}
\usepackage{pos}
\usepackage{lipsum} 

\title{Study of an Isolated Double-pulse Cosmic Ray Candidate Recorded with the Askaryan Radio Array}

\ShortTitle{Study of a Double-pulse Cosmic Ray Candidate Recorded with the ARA}

\author*[a]{Shoukat Ali} 
\author[a]{Dave Z. Besson}
\onbehalf{for the ARA Collaboration \\{\normalsize \normalfont(a complete list of authors can be found at the end of the proceedings)}\\}

\affiliation[a]{Department of Physics and Astronomy, University of Kansas,\\
  1450 Jayhawk Blvd, Lawerence 66045, USA}

\emailAdd{shoukat@ku.edu}
\emailAdd{zedlam@ku.edu}

\abstract{The radio-frequency emissions produced by particle showers on Earth, resulting from cosmic rays (CRs) and ultra-high energy neutrinos (UHE-$\nu$) originating from astrophysical sources share significant similarities, enabling radio detectors initially designed for UHE-$\nu$ searches to also study CRs. The Askaryan Radio Array (ARA), an experiment currently operating within the ice at the South Pole, is primarily designed to detect UHE-$\nu$s. To date, ARA has deployed five stations, with each station equipped with antennas installed at depths up to 200 meters in the ice.
Data recorded by ARA Station-2 (A2) suggest a potential CR origin for a subset of events identified in a UHE-$\nu$ search. This subset includes a double-pulse event potentially from a downward propagating CR-induced air shower, with in-air geomagnetic emission followed by in-ice Askaryan emission producing the two pulses. A detailed investigation of this CR candidate event using comprehensive simulations has been conducted with the goal of identifying the parameters of a CR-induced air shower that best match the experimentally observed quantities. We simulate predicted CR signals in ARA by combining an impacting CR shower simulation framework (FAERIE) with a realistic detector simulation (AraSim). \\
We determine the event topology based on the vertex reconstruction of both the putative geomagnetic and Askaryan signals. After inferring the event geometry, we show that the simulation matches the observed time structure of the event (channel-by-channel relative signal arrival times) for the recorded event.
}
\ConferenceLogo{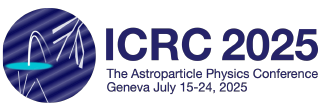}

\FullConference{39th International Cosmic Ray Conference (ICRC2025)\\
 15–24 July 2025\\
Geneva, Switzerland\\}

\begin{document}

\maketitle

\section{Introduction}\label{sec1}
Cosmic rays (CRs) are among the most important messengers in multi-messenger astrophysics, offering a unique window into the universe. In particular, ultra-high energy cosmic rays (UHE-CRs) have the potential to uncover new physics through their particle interactions. In addition to conventional techniques for CR detection, radio detection provides a valuable complementary method. This approach began with the pioneering observation of radio pulses from extensive air showers by Jelley and Fruin in 1965~\cite{j65}. Modern radio detection techniques have since advanced and are now employed in radio neutrino detectors. Although primarily designed to search for UHE-$\nu$s, these detectors are also sensitive to radio signals produced by CR-induced air showers, which arise through both geomagnetic and Askaryan emission mechanisms.\\
When a CR enters the Earth's atmosphere, it interacts inelastically with atmospheric nuclei, which subsequently develops a cascade of particles (predominantly electrons, positrons, muons, pions, and photons), called an `extensive air shower'. `Geomagnetic emission' arises due to the deflection of charged particles in the Earth’s magnetic field via the Lorentz force, which induces a transverse current in the shower front. The time variation of this current produces linearly polarized radiation, with the polarization direction oriented perpendicular to the geomagnetic Lorentz force vector. This mechanism accounts for approximately $90\%$ of the total observed in-air radio signal. At higher energies, the core of CR-induced air showers can penetrate the ice and continue to evolve, developing a net negative charge excess as they propagate through the upper 10 meters of the ice sheet. This moving charge, traveling at velocities close to the speed of light in the medium produces coherent radio emission via the Askaryan effect, where the electric field is polarized towards the center of the shower core \cite{Askaryan}. In-ice radio detectors are capable of detecting both geomagnetic and Askaryan components of the radio signal, which can appear as distinct peaks in the recorded waveforms\cite{Javaid2012}\cite{deVries2016}. Due to differences in the propagation paths, the geomagnetic signal typically arrives earlier than the Askaryan signal. The time delay between the two components depends on several factors, including the detector's position, the shower’s impact location, the energy of the primary cosmic ray, and its zenith and azimuth angles, collectively defining the event topology. \\

The Askaryan Radio Array (ARA) is a radio neutrino detector located in the Antarctic ice at the geographic South Pole, designed to detect UHE-$\nu$. To date, five ARA stations have been deployed, including one equipped with a phased array trigger system to enhance sensitivity to low signal-to-noise ratio(SNR) events. Each station consists of six deep boreholes (strings), four of which are dedicated to signal reception and two to calibration. Each signal string contains four broadband receiving antennas: two vertically polarized (VPol) and two horizontally polarized (HPol). A typical layout places a VPol antenna at the lowest depth, followed by an HPol antenna 2 meters above, with a second VPol-HPol pair positioned 20 meters higher. The two calibration strings, located approximately 40–50 meters from the station center, each have pulsers with one VPol and one HPol antenna for detector calibration. The receiving antennas are sensitive over a frequency range of 130–850 MHz\cite{Allison2016}\cite{mh23}. For each triggered event, the ARA electronics system digitizes and records sixteen $\sim$500-ns long voltage waveforms, corresponding to the sixteen receiving antennas across the four signal strings.

ARA station-2 (A2) recorded a double pulse event identified in an eight station-year UHE-$\nu$ search \cite{brianthesis}, and subsequently shown at conferences\cite{ammyTalk}\cite{shoukatTalk}. This UHE-$\nu$ search analysis implemented a rigorous event-level thermal noise rejection algorithm, complemented by a comprehensive set of selection criteria designed to suppress all known sources of anthropogenic interference for ARA. \\

In this work, we provide a Monte Carlo-based interpretation of this double-pulse event and demonstrate its consistency with the in-air geomagnetic and in-ice Askaryan emissions arising from a CR-induced air shower. We note that this event was also uniquely identified in an independent dedicated search for non-anthropogenic high-SNR double-pulse events in two A2 station-years of data. For that search, events were selected which had two impulsive excursions 1) separated by at least 40 ns, 2) in at least two A2 channels, and 3) for which the summed waveform power in both the VPol and also the HPol channels was at least 4 times greater than the average measured waveform power for forced triggers. Those criteria isolated two surviving events in the 2015-2016 data samples, one of which was an event for which two calibration pulser signals were (spuriously) emitted nearly simultaneously, and the other was the geomagnetic-Askaryan candidate event we investigate here. 

\section{Event Analysis}\label{sec2}
\begin{figure}[h]
    \includegraphics[width=0.83\linewidth, height=6.3cm]{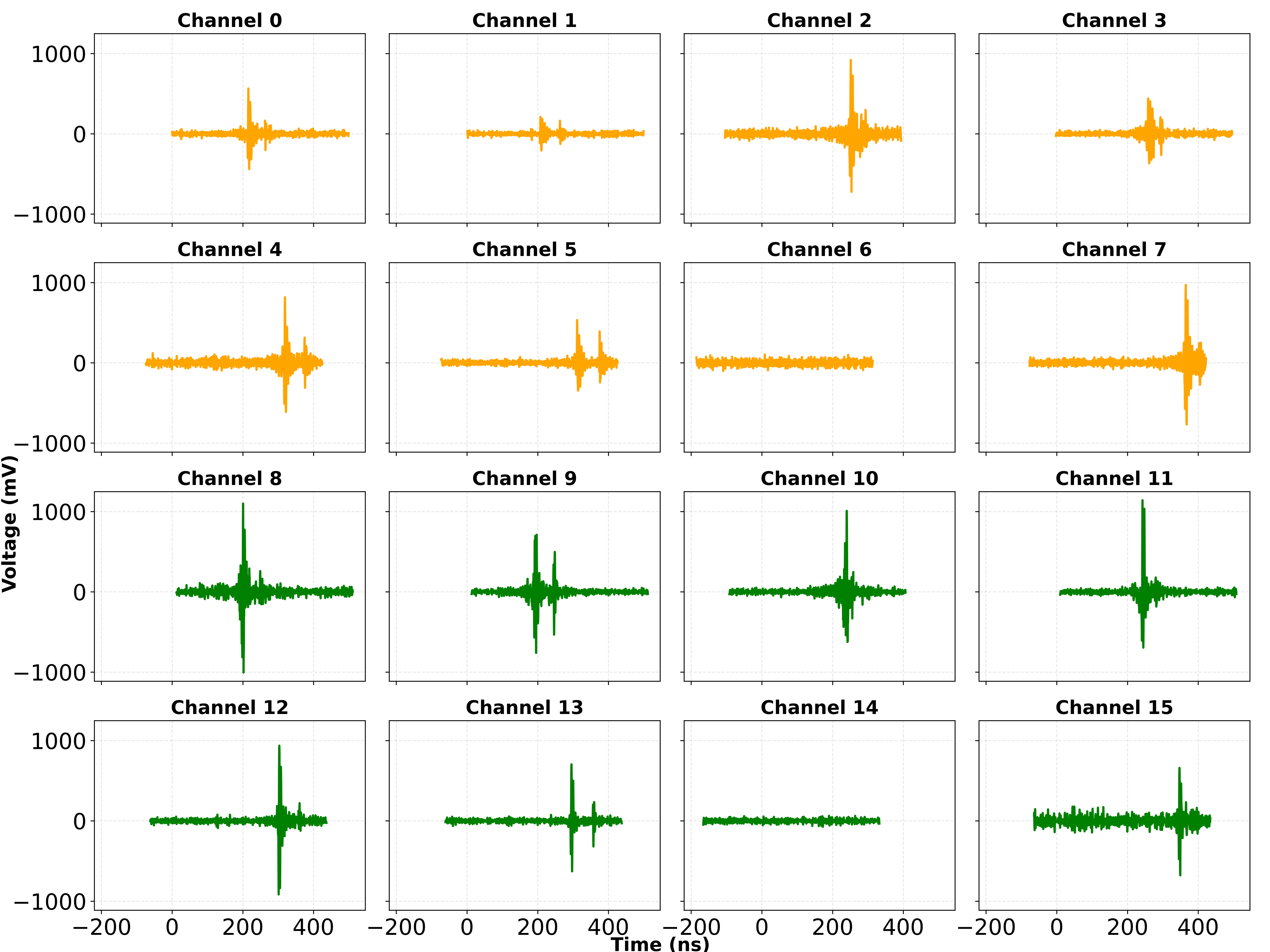}
    \caption{Waveforms of the CR candidate event for sixteen channels recorded by A2 in 2018. The top two rows (orange) correspond to VPol channels, and the bottom two rows (blue) represent HPol channels.}
    \label{fig:ev_wf}
\end{figure}

The de-dispersed waveforms of the CR candidate event are shown in Figure~\ref{fig:ev_wf}. These waveforms have been corrected for the frequency-dependent phase response of the signal chain using the standard ARA de-dispersion tool. The event exhibits distinct double pulses in channels 0, 1, 4, 5, 7, 9, 12, and 13. Except for channels 6 and 8, the first pulse is clearly observed across all channels. The variation in the channel-by-channel time delay between the two pulses suggests that they arise from two different emission vertices, rather than a single source of double-pulses, which would yield identical time delays between the two. Since the geomagnetic signal typically arrives before the Askaryan signal, we interpret the first and second pulses as putative geomagnetic and Askaryan signals, respectively. \\

For inclined CR-induced air showers, the geomagnetic radio emission is mostly horizontally polarized because the Earth's magnetic field at the South Pole is nearly vertical. In contrast, the conical Askaryan Cherenkov signal, projected onto the receiver antennas, may include both polarization components. As a result, Hpol antennas are expected to receive more power from geomagnetic signals than VPol antennas. This difference can be quantified using the total received event power ratio HPol/VPol, which tends to be larger for geomagnetic signals than Askaryan. For the event analyzed in this study, we calculated the polarization power ratio, defined as ($ \frac{\sum_{ch=0}^{16} A_{HPol}}{\sum_{ch=0}^{16} A_{VPol}}$) for each channel using the following expression \ref{eq:apulse}.
\begin{equation}
    A_{pulse} = \sqrt{A_{env}^2-V_{\rm rms}^2}
    \label{eq:apulse}
\end{equation}
Here, $A_{env}$ denotes the peak amplitude of the Hilbert envelope of the waveform, while the RMS is estimated from a pre-signal region to account for thermal noise contributions. After summing over all channels, this ratio is found to be 1.513 for the first pulse and 0.922 for the second pulse, consistent with the assumption of geomagnetic followed by an Askaryan emission. \\

Vertex reconstruction of the two pulses was performed in the ARA Station-Centric coordinate system, for which the origin is defined at a depth of 179.9 meters in the ice, and the z-axis points upward. The zenith angle is measured relative to the z-axis, while the azimuthal angle is measured counter-clockwise from the x-axis, which is aligned with the direction of the local ice flow. To isolate the sources, the two pulses were separated and reconstructed independently. The standard ARA interferometric reconstruction algorithm was used to determine the arrival direction of each pulse and cross-validated with a simpler triangulation reconstruction algorithm \cite{AraVtx}\cite{Interferometry}. The resulting reconstructed directions are shown in Figure \ref{fig:sim_rec_dt}. 
For the channels exhibiting clear double pulses, the measured time delays between the two pulses are shown in Figure \ref{fig:sim_rec_dt}.

\section{Simulations}\label{sec3}
To understand the origin of the two pulses seen in the A2 event, we use FAERIE (Framework for the Simulation of Air Shower Emission of Radio for In-Ice Experiments), a recently developed Monte Carlo simulation tool. FAERIE\cite{faerie} simulates the development of CR-induced air showers in the atmosphere, their propagation and evolution through air and ice, and the transmission of the resulting radio signals via both the geomagnetic and Askaryan emission mechanisms to the in-ice antennas .
FAERIE integrates CORSIKA 7.7500\cite{corsika} (for simulating the propagation of the air shower through the atmosphere and including an altitude-dependent index of refraction model for air), a modified version of COREAS (for calculating the radio emission generated by the shower using the end-point formalism to determine the electric field at specified antenna positions in the ice), and GEANT4 10.5\cite{GEANT4} to model the generation of the in-ice particle shower and propagation of the shower core through the variable-density ice\cite{kennyIcemodel}. The transmission of COREAS radio signals from the particle shower to the in-ice antennas is simulated using a ray-tracing approach, which accounts for the variation in the index of refraction as the signal travels through different media such as air and ice. The modified COREAS module uses a local geomagnetic coordinate system to define the geometry of the cosmic-ray-induced shower and the positions of the receiving antennas. This setup ensures that both the direction and polarization of the emitted radio signals are accurately modeled as they propagate toward the detector. In the last step, the
radio emission produced by the particle cascade using the endpoint formalism\cite{endpoint} is calculated and propagated to the in-ice antennas. These time-domain electric field traces simulated by FAERIE are processed through AraSim to model the detector response, resulting in the corresponding voltage traces.

\section{Results}\label{sec4}
The shower zenith angle $\theta_s$, azimuth $\phi_s$, and surface impact parameter $d_s$ were determined using reconstruction outputs and first-principles modeling of a CR-induced air shower with the A2 detector geometry. These quantities are derived from the reconstructed pulse parameters as follows:
\begin{subequations}
\begin{align}
    \theta_s &= \sin^{-1}\left(\frac{n_{\text{ice}} \cdot \sin(\theta_{\text{rec, 1stpulse}})}{n_{\text{air}}}\right) \tag{1} \\
    \varphi_s &= 180^\circ - \phi_{\text{rec}} \tag{2} \\
    d_s &= (d_{\text{center}} - d_{\text{max}}) \cdot \tan(\theta_{\text{rec, 2ndpulse}}) + d_{\text{max}} \cdot \tan(\theta_{\text{rec, 2ndpulse}} + \theta_c) \tag{3}
\end{align}
\end{subequations}

\begin{figure}[htb!]
    \centering
    \begin{minipage}[b]{0.47\linewidth}
        \centering
        \includegraphics[width=\linewidth]{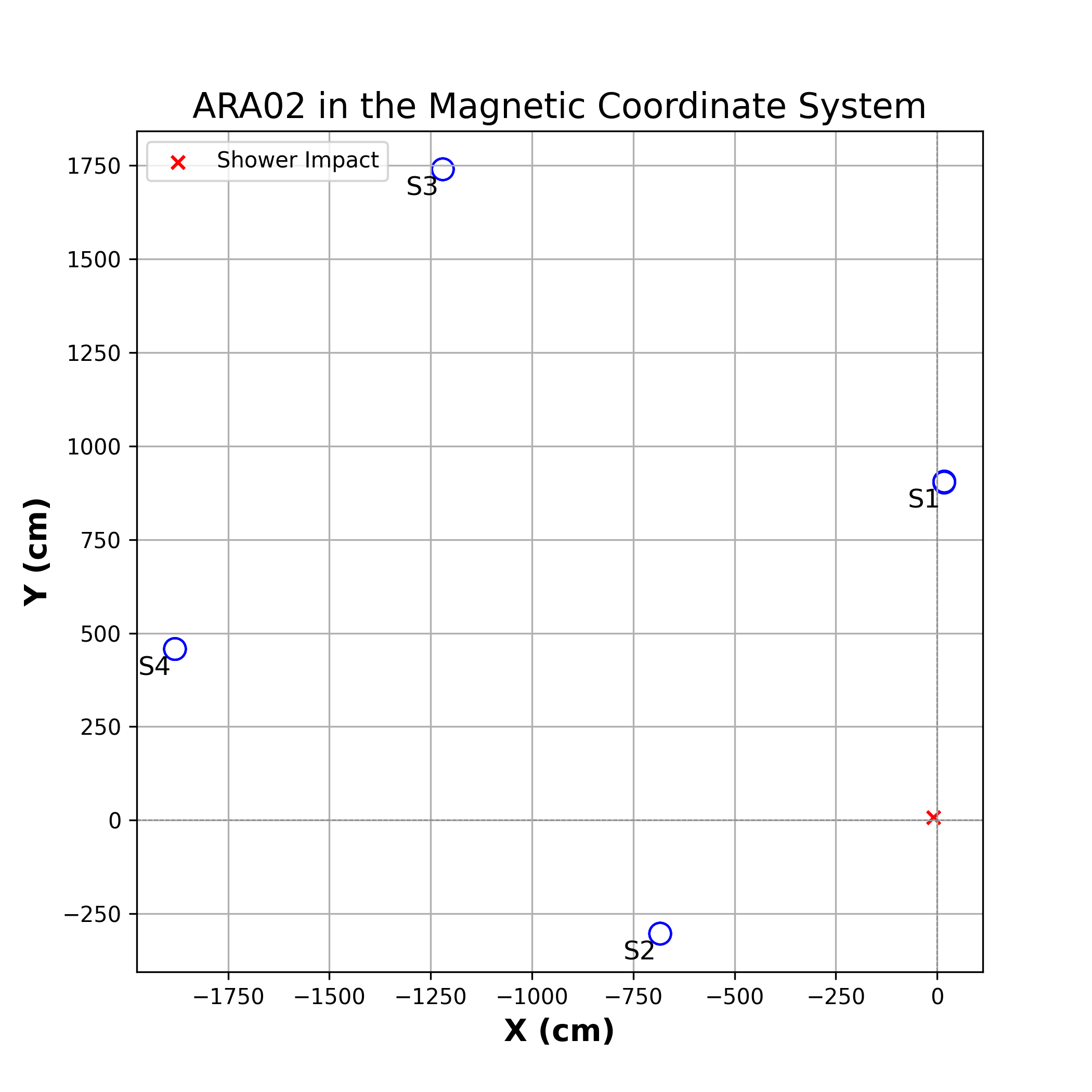}
        \label{fig:ev_top_a}
    \end{minipage}
    \hfill
    \begin{minipage}[b]{0.48\linewidth}
        \centering
        \includegraphics[width=\linewidth, height=6.7cm]{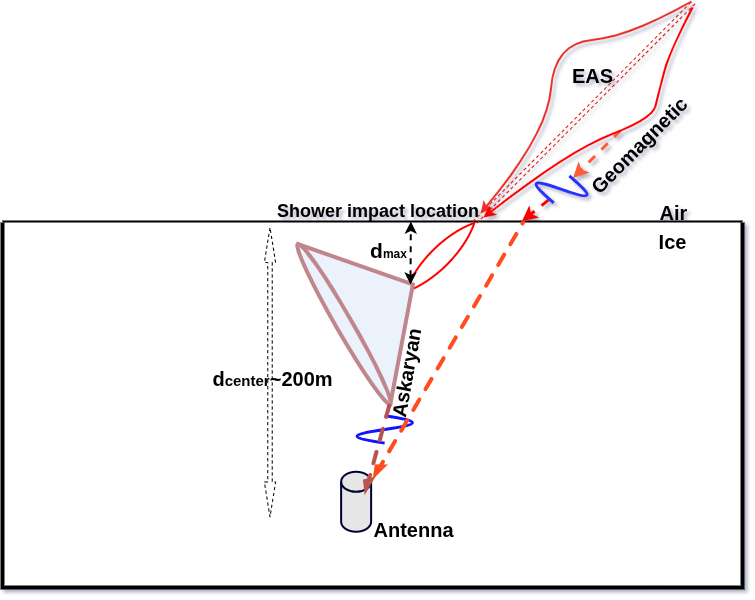}
        \label{fig:ev_top_b}
    \end{minipage}
    \caption{(Left): A2 strings (S1, S2, S3, S4) and the shower impact location shown relative to each other. (Right): Event topology corresponding to the presented results.}
    \label{fig:event_top}
\end{figure}

\begin{figure}[htb!]
    \centering
    \begin{minipage}[b]{0.48\linewidth}
        \centering
        \includegraphics[width=\linewidth]{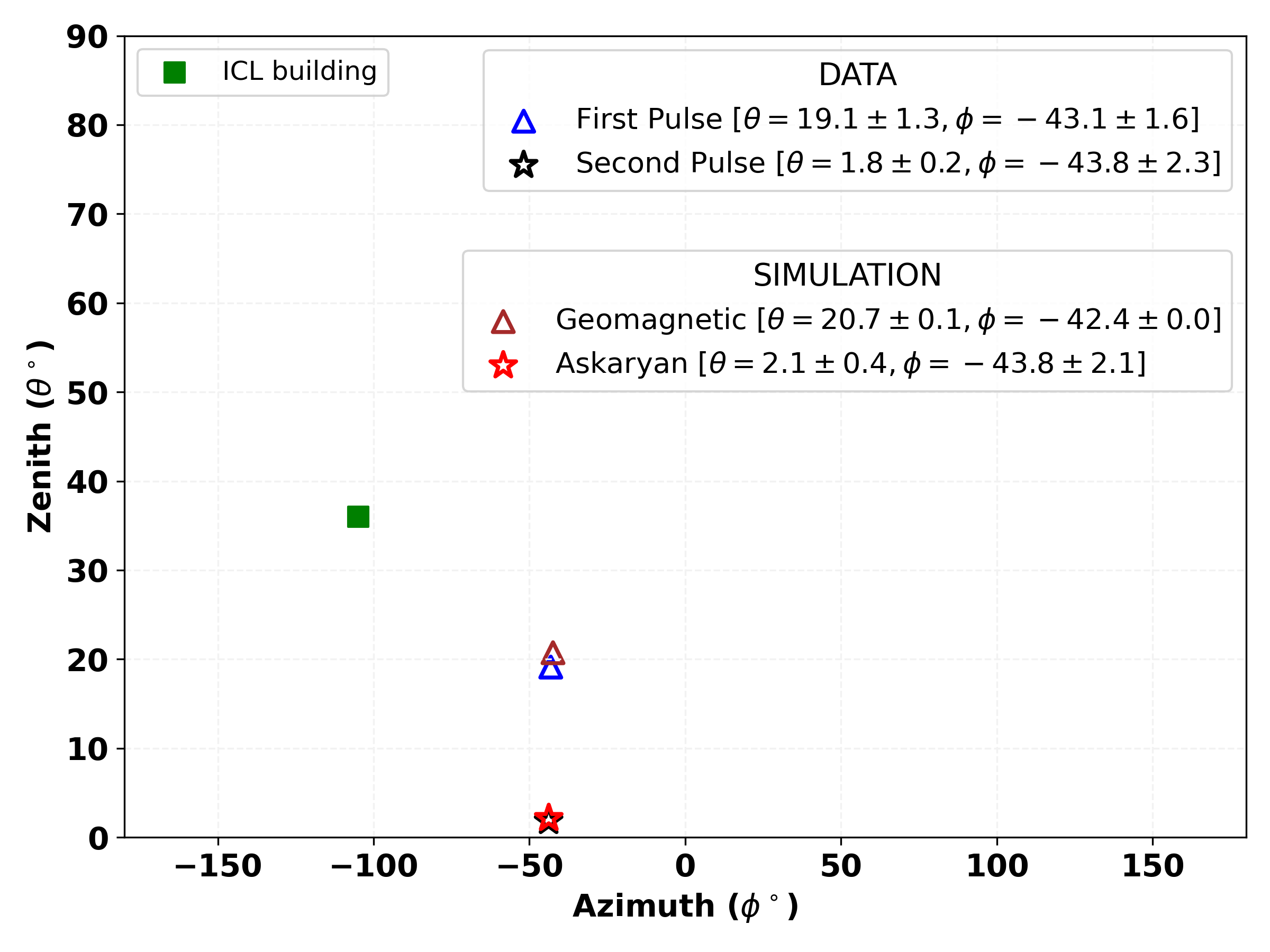}
        \label{fig:ev_top_a}
    \end{minipage}
    \hfill
    \begin{minipage}[b]{0.48\linewidth}
        \centering
        \includegraphics[width=\linewidth]{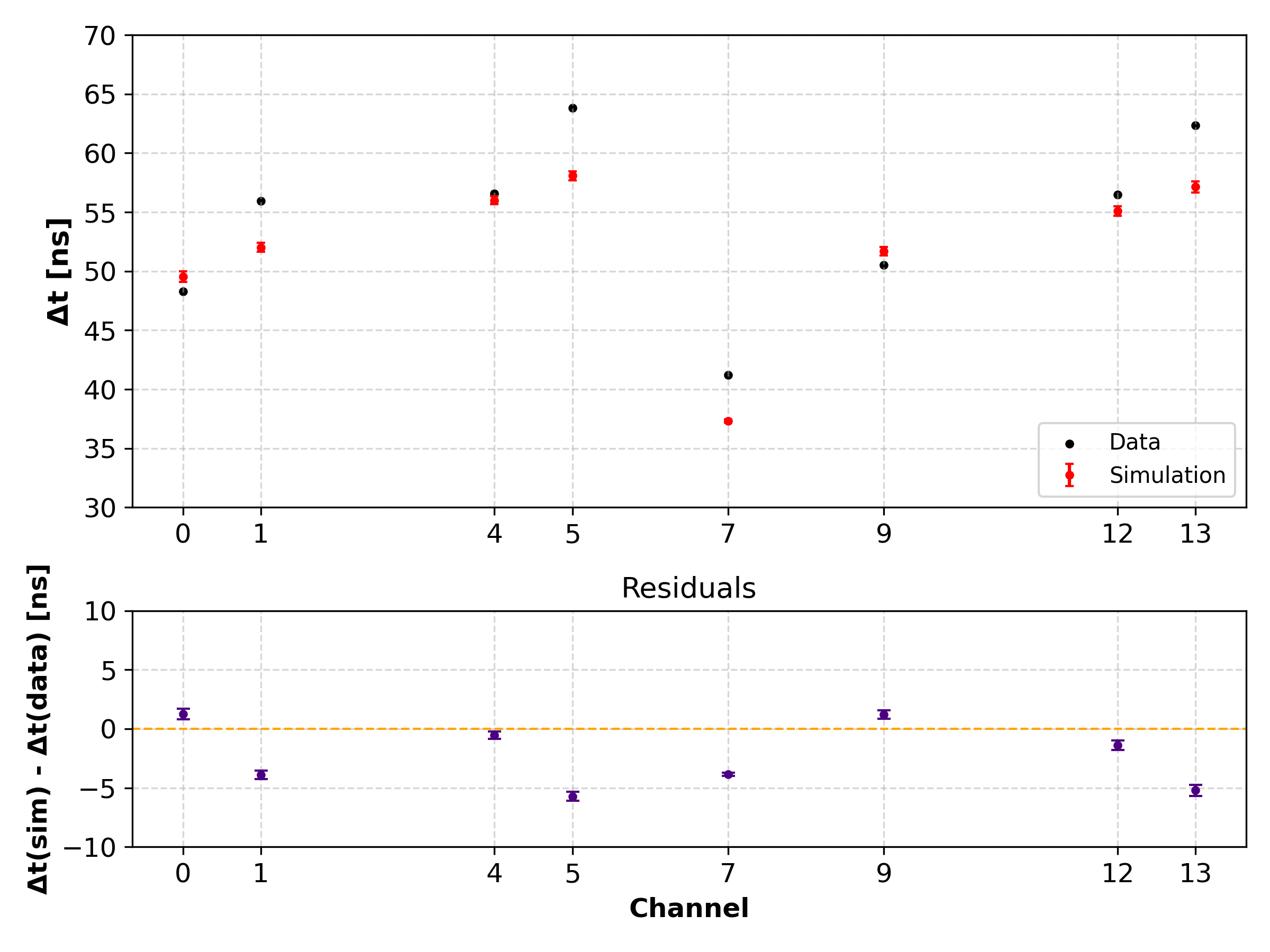}
        \label{fig:ev_top_b}
    \end{minipage}
    \caption{(Left) Vertex direction of the first and second pulses from the candidate event compared with geomagnetic and Askaryan signals from simulated CR-induced showers. The plot shows average values and uncertainties, obtained using two independent reconstruction algorithms of ARA. (Right) Measured vs. simulated time delays for preferred signals (top), with residuals (lower panel). The statistical uncertainties in the time delays were determined based on ten simulated shower events.}
    \label{fig:sim_rec_dt}
\end{figure}

The calculations assume a bulk refractive index for ice of $n_{\text{ice}} = 1.78$, an in-ice shower maximum depth of $d_{\text{max}} = 6\,\mathrm{m}$, and a Cherenkov angle of $\theta_c = 43^\circ$, based on the local refractive index. The inferred event topology, constrained by these parameters, along with antenna positions mapped into the magnetic coordinate system, is illustrated in Figure~\ref{fig:event_top}. The magnetic coordinate system is a rectangular coordinate system, in which the x-axis and y-axis point to magnetic north and west, respectively, with the z-axis pointing perpendicular to the x-y plane. 
Assuming a proton as the primary CR and setting the primary particle energy to 10 PeV, CR showers for the preferred event topology were analyzed. Vertex reconstruction was performed for both the geomagnetic and Askaryan components of the radio signal received by the A2 antennas. The reconstructed vertices are in good agreement with those obtained from the first and second pulses of the observed candidate event, as shown in Figure~\ref{fig:sim_rec_dt}(Left). After performing reconstruction by isolating the two pulses, the inferred reconstructed event topology also prescribes time delays between the arrival of the geomagnetic and Askaryan signals in each channel. A comparison between the calculated time delays from the simulated showers and those observed in the data is also presented in Figure~\ref{fig:sim_rec_dt}(Right), demonstrating consistency with the simulation-based expectations. In contrast to timing (and as recognized by a companion ARA submission to this conference\cite{cr_a5pos}), obtaining amplitude agreement between simulation and data is complicated by i) large (of order unity) intrinsic shower-to-shower fluctuations owing to the stochastic nature of the EAS development, and ii) difficulties in precise propagation of the collimated ($\sim1^\circ$) Askaryan signal through the variable (and fluctuating) density firn, to the 200-m deep receiver antennas.

\section{Conclusion}\label{sec5}
The A2 CR candidate event with a double-pulse feature was analyzed in detail using microscopic simulation. An event topology was deduced for which the geomagnetic signal arrives at the antenna before the Askaryan signal, leading to measurable time delays between the two pulses in each channel.\\

To test this interpretation, CR-induced air showers were simulated using the A2 station geometry. The reconstructed vertex direction and the time delays between pulses in different antenna channels exhibit good agreement with the observed data, making the CR-hypothesis a favored interpretation for the recorded double-pulse event.\\
As a next step, we will calculate the expected polarization angles of the geomagnetic and Askaryan emissions from the event geometry and compare them to A2 data measurements. We will also extract the Hpol/Vpol power ratio from simulated showers to quantify each mechanism's contribution. Combining these analyses will provide a thorough validation of the polarization behavior, strengthening the interpretation of the double-pulse event's cosmic-ray origin.
{\footnotesize  
\setlength{\itemsep}{0pt}
\setlength{\parsep}{0pt}
\setlength{\parskip}{0pt}

}
\clearpage

\section*{Full Author List: ARA Collaboration (June 30, 2025)}

\noindent
N.~Alden\textsuperscript{1}, 
S.~Ali\textsuperscript{2}, 
P.~Allison\textsuperscript{3}, 
S.~Archambault\textsuperscript{4}, 
J.J.~Beatty\textsuperscript{3}, 
D.Z.~Besson\textsuperscript{2}, 
A.~Bishop\textsuperscript{5}, 
P.~Chen\textsuperscript{6}, 
Y.C.~Chen\textsuperscript{6}, 
Y.-C.~Chen\textsuperscript{6}, 
S.~Chiche\textsuperscript{7}, 
B.A.~Clark\textsuperscript{8}, 
A.~Connolly\textsuperscript{3}, 
K.~Couberly\textsuperscript{2}, 
L.~Cremonesi\textsuperscript{9}, 
A.~Cummings\textsuperscript{10,11,12}, 
P.~Dasgupta\textsuperscript{3}, 
R.~Debolt\textsuperscript{3}, 
S.~de~Kockere\textsuperscript{13}, 
K.D.~de~Vries\textsuperscript{13}, 
C.~Deaconu\textsuperscript{1}, 
M.A.~DuVernois\textsuperscript{5}, 
J.~Flaherty\textsuperscript{3}, 
E.~Friedman\textsuperscript{8}, 
R.~Gaior\textsuperscript{4}, 
P.~Giri\textsuperscript{14}, 
J.~Hanson\textsuperscript{15}, 
N.~Harty\textsuperscript{16}, 
K.D.~Hoffman\textsuperscript{8}, 
M.-H.~Huang\textsuperscript{6,17}, 
K.~Hughes\textsuperscript{3}, 
A.~Ishihara\textsuperscript{4}, 
A.~Karle\textsuperscript{5}, 
J.L.~Kelley\textsuperscript{5}, 
K.-C.~Kim\textsuperscript{8}, 
M.-C.~Kim\textsuperscript{4}, 
I.~Kravchenko\textsuperscript{14}, 
R.~Krebs\textsuperscript{10,11}, 
C.Y.~Kuo\textsuperscript{6}, 
K.~Kurusu\textsuperscript{4}, 
U.A.~Latif\textsuperscript{13}, 
C.H.~Liu\textsuperscript{14}, 
T.C.~Liu\textsuperscript{6,18}, 
W.~Luszczak\textsuperscript{3}, 
A.~Machtay\textsuperscript{3}, 
K.~Mase\textsuperscript{4}, 
M.S.~Muzio\textsuperscript{5,10,11,12}, 
J.~Nam\textsuperscript{6}, 
R.J.~Nichol\textsuperscript{9}, 
A.~Novikov\textsuperscript{16}, 
A.~Nozdrina\textsuperscript{3}, 
E.~Oberla\textsuperscript{1}, 
C.W.~Pai\textsuperscript{6}, 
Y.~Pan\textsuperscript{16}, 
C.~Pfendner\textsuperscript{19}, 
N.~Punsuebsay\textsuperscript{16}, 
J.~Roth\textsuperscript{16}, 
A.~Salcedo-Gomez\textsuperscript{3}, 
D.~Seckel\textsuperscript{16}, 
M.F.H.~Seikh\textsuperscript{2}, 
Y.-S.~Shiao\textsuperscript{6,20}, 
S.C.~Su\textsuperscript{6}, 
S.~Toscano\textsuperscript{7}, 
J.~Torres\textsuperscript{3}, 
J.~Touart\textsuperscript{8}, 
N.~van~Eijndhoven\textsuperscript{13}, 
A.~Vieregg\textsuperscript{1}, 
M.~Vilarino~Fostier\textsuperscript{7}, 
M.-Z.~Wang\textsuperscript{6}, 
S.-H.~Wang\textsuperscript{6}, 
P.~Windischhofer\textsuperscript{1}, 
S.A.~Wissel\textsuperscript{10,11,12}, 
C.~Xie\textsuperscript{9}, 
S.~Yoshida\textsuperscript{4}, 
R.~Young\textsuperscript{2}
\\
\\
\textsuperscript{1} Dept. of Physics, Enrico Fermi Institute, Kavli Institute for Cosmological Physics, University of Chicago, Chicago, IL 60637\\
\textsuperscript{2} Dept. of Physics and Astronomy, University of Kansas, Lawrence, KS 66045\\
\textsuperscript{3} Dept. of Physics, Center for Cosmology and AstroParticle Physics, The Ohio State University, Columbus, OH 43210\\
\textsuperscript{4} Dept. of Physics, Chiba University, Chiba, Japan\\
\textsuperscript{5} Dept. of Physics, University of Wisconsin-Madison, Madison,  WI 53706\\
\textsuperscript{6} Dept. of Physics, Grad. Inst. of Astrophys., Leung Center for Cosmology and Particle Astrophysics, National Taiwan University, Taipei, Taiwan\\
\textsuperscript{7} Universite Libre de Bruxelles, Science Faculty CP230, B-1050 Brussels, Belgium\\
\textsuperscript{8} Dept. of Physics, University of Maryland, College Park, MD 20742\\
\textsuperscript{9} Dept. of Physics and Astronomy, University College London, London, United Kingdom\\
\textsuperscript{10} Center for Multi-Messenger Astrophysics, Institute for Gravitation and the Cosmos, Pennsylvania State University, University Park, PA 16802\\
\textsuperscript{11} Dept. of Physics, Pennsylvania State University, University Park, PA 16802\\
\textsuperscript{12} Dept. of Astronomy and Astrophysics, Pennsylvania State University, University Park, PA 16802\\
\textsuperscript{13} Vrije Universiteit Brussel, Brussels, Belgium\\
\textsuperscript{14} Dept. of Physics and Astronomy, University of Nebraska, Lincoln, Nebraska 68588\\
\textsuperscript{15} Dept. Physics and Astronomy, Whittier College, Whittier, CA 90602\\
\textsuperscript{16} Dept. of Physics, University of Delaware, Newark, DE 19716\\
\textsuperscript{17} Dept. of Energy Engineering, National United University, Miaoli, Taiwan\\
\textsuperscript{18} Dept. of Applied Physics, National Pingtung University, Pingtung City, Pingtung County 900393, Taiwan\\
\textsuperscript{19} Dept. of Physics and Astronomy, Denison University, Granville, Ohio 43023\\
\textsuperscript{20} National Nano Device Laboratories, Hsinchu 300, Taiwan\\

\section*{Acknowledgements}

\noindent
The ARA Collaboration is grateful to support from the National Science Foundation through Award 2013134.
The ARA Collaboration
designed, constructed, and now operates the ARA detectors. We would like to thank IceCube, and specifically the winterovers for the support in operating the
detector. Data processing and calibration, Monte Carlo
simulations of the detector and of theoretical models
and data analyses were performed by a large number
of collaboration members, who also discussed and approved the scientific results presented here. We are
thankful to Antarctic Support Contractor staff, a Leidos unit 
for field support and enabling our work on the harshest continent. We thank the National Science Foundation (NSF) Office of Polar Programs and
Physics Division for funding support. We further thank
the Taiwan National Science Councils Vanguard Program NSC 92-2628-M-002-09 and the Belgian F.R.S.-
FNRS Grant 4.4508.01 and FWO. 
K. Hughes thanks the NSF for
support through the Graduate Research Fellowship Program Award DGE-1746045. A. Connolly thanks the NSF for
Award 1806923 and 2209588, and also acknowledges the Ohio Supercomputer Center. S. A. Wissel thanks the NSF for support through CAREER Award 2033500.
A. Vieregg thanks the Sloan Foundation and the Research Corporation for Science Advancement, the Research Computing Center and the Kavli Institute for Cosmological Physics at the University of Chicago for the resources they provided. R. Nichol thanks the Leverhulme
Trust for their support. K.D. de Vries is supported by
European Research Council under the European Unions
Horizon research and innovation program (grant agreement 763 No 805486). D. Besson, I. Kravchenko, and D. Seckel thank the NSF for support through the IceCube EPSCoR Initiative (Award ID 2019597). M.S. Muzio thanks the NSF for support through the MPS-Ascend Postdoctoral Fellowship under Award 2138121. A. Bishop thanks the Belgian American Education Foundation for their Graduate Fellowship support.

\end{document}